\documentclass[twocolumn,showpacs,preprintnumbers,floatfix,
               amsmath,amssymb]{revtex4}
\usepackage{longtable}
\usepackage{graphics}
\usepackage{epsfig}
\usepackage{bm}

\begin{document}
\preprint{FT-053}

\title{Color screening in a constituent 
       quark model of hadronic matter}

\author{G. Toledo S\'anchez}\email{toledo@fisica.unam.mx}
\affiliation{Instituto de F\'{\i}sica UNAM, A.~P.~20-364 
             M\'exico 01000 D.~F.~M\'exico}
\author{J. Piekarewicz}\email{jorgep@csit.fsu.edu}
\affiliation{Department of Physics,
             Florida State University, Tallahassee, FL 32306-4350}

\pacs{24.85.+p, 24.10.Lx, 25.75.Nq, 12.39.Jh}

\date{\today}

\begin{abstract}
The effect of color screening on the formation of a heavy
quark-antiquark ($Q\bar{Q}$) bound state---such as the $J/\psi$
meson---is studied using a constituent-quark model. The response of
the nuclear medium to the addition of two color charges is simulated
directly in terms of its quark constituents via a string-flip
potential that allows for quark confinement within hadrons yet enables
the hadrons to separate without generating unphysical long-range
forces.  Medium modifications to the properties of the heavy meson,
such as its energy and its mean-square radius, are extracted by
solving Schr\"odinger's equation for the $Q\bar{Q}$ pair in the
presence of a (screened) density-dependent potential. The density
dependence of the heavy-quark potential is in qualitative agreement
with earlier studies of its temperature dependence extracted from
lattice calculations at finite temperature. In the present model it is
confirmed that abrupt changes in the properties of the $J/\psi$-meson
in the hadronic medium ({\it plasma}), correlate strongly with the
deconfining phase transition.
\end{abstract}
\maketitle
%
\section{Introduction}
\label{introduction}

The quest for {\it novel states of matter} remains a central theme in
physics and one that spans all of its subfields. The discovery of
high-temperature superconductors and the experimental realization of
Bose-Einstein condensation are some recent examples. The widespread
impact of these discoveries in the areas of condensed-matter physics
and atomic physics is well documented in the literature. At the
interface between nuclear and particle physics---but with impact in
cosmology and astrophysics---is the search for the quark-gluon plasma
(QGP), a novel state of matter that may have existed at the dawn of
the Universe and that may still exist today in the dense environments
of exotic stars. The quark-gluon plasma, a {\it deconfined} state of
quarks and gluons, is predicted from QCD to be attained at high
temperatures and/or high baryon densities. A variety
of experiments have been devoted to produce the QGP in the laboratory.
These efforts started at the AGS in Brookhaven and at the SPS in CERN,
and now continue at the relativistic-heavy-ion collider (RHIC), and
should culminate with the construction of the large-hadron-collider
(LHC) at CERN~\cite{Bay02_NPA698}. By colliding extremely energetic
heavy ions, the aim of these experiments is to create a region of such
high-energy density (or high temperature) that quarks and gluons will
become deconfined. While current experimental facilities (primarily
RHIC) may have already created the QGP~\cite{Adl04_PRC69}, a great 
challenge remains: how to identify clearly and unambiguously the 
production of such coveted state?

Spatial observations complement terrestrial searches for the QGP.  The
advent of sophisticated telescopes operating at a variety of
wavelengths have turned neutron stars from theoretical curiosities
into powerful diagnostic tools. The observation of an anomalous
mass-radius relation and/or an enhanced cooling in neutron stars, may
provide strong evidence in support of strange
stars~\cite{Xu02_APJ570}, quark stars~\cite{Dra02_APJ572}, or neutron
stars with exotic cores. These cores may contain novel states of
matter, such as meson condensates, strange-quark matter, and/or color
superconductors~\cite{Bay79_AA17,Wit84_PRD30,Alf98_PLB422}.

In the case of terrestrial searches for the QGP, there is a set of
complementary observables which provide information at different
stages of the experiment. For example, leptons and photons produced at
the center of the collision are expected to carry valuable information
concerning the earliest stages of the collisions, while the relative
abundance of light-flavor hadrons should reflect the freeze-out
stage. Evidence for the QGP is also imprinted in the response of the
medium to the formation and propagation of a heavy quark-antiquark
pair, where Debye screening of the bound state ({\it e.g.,} $J/\psi$)
is expected~\cite{Mat86_PLB178,Mat03_PTP151}. In this case it is
important to study the possible {\it dissolution} of the bound state
and the evolution of the confining potential with baryon and/or energy
density. While the quest for these experimental signatures will 
undoubtedly continue, convincing arguments in favor of the discovery 
of a strongly-coupled QGP at RHIC have recently been 
made~\cite{Gyu04_nt03032}. This persuasive study, containing a host 
of valuable references, uses existing observables---such as bulk 
collective flow and jet quenching---to justify the claim. 

In the present work we study the efficiency of the nuclear medium in
screening the color charges of a heavy quark-antiquark pair. That is, 
we study color screening as a function of the baryon density rather 
than as a function of temperature. This distinction is important as 
our $T\!=\!0$ formalism is unable to shed light on finite-temperature 
calculations, which themselves appear to be in a state of flux. 
Calculations based on potential models predict the $J/\psi$-meson to 
dissolve at a temperature of 
$T\!\simeq\!1.1~\!T_{\rm c}$~\cite{Dig01_PRD64} (with $T_{\rm c}$
being the critical temperature for the deconfining phase transition). 
However, recent lattice results seem to suggest that the
$J/\psi$-meson survives up to a temperature of
$T\!\simeq\!1.5~\!T_{\rm c}$~\cite{Dat03_NPB119,Dat03_HL12037}. 
Moreover, the originally strong suppression of the $J/\psi$-meson 
predicted by Matsui and Satz has been recently put into question. 
Indeed, it has been argued that this elusive suppression 
may be offset (at least in part) by statistical recombination of 
$J/\psi$-mesons from the initially produced $c\bar{c}$ 
pairs~\cite{Bra00_PLB490,The01_PRC63,And03_PLB571,The04_JPG30}. 
We offer no new insights into these important problems. Rather, we 
consider a simple model that dynamically interpolates between a 
hadron-based description at low-density and a quark-based description 
at high-density to study the dynamics of the $J/\psi$-meson. In
essence, we aim at computing a heavy-quark potential as a function
of baryon density to complement (and compare to) lattice 
studies~\cite{Kar01_NPA605,Ume03_NPA721,Rap04_JPG30} that, while 
successful at finite-temperature, are unable to simulate hadronic 
systems at finite baryon density.

To correlate the nuclear-to-quark-matter transition to the in-medium
properties of the $J/\psi$ meson, a {\it string-flip model} is
used~\cite{Len86_AP170}. Although we will refer explicitly to the 
$J/\psi$ meson henceforth, our approach is applicable to any heavy meson (or baryon) added to the hadronic medium. The
string-flip model belongs to a class of constituent-quark models that
incorporates many of the phenomenological features of quantum
chromodynamics (QCD). The underlying degrees of freedom are
constituent quarks that interact via confining strings (or {\it flux
tubes}) that arrange themselves instantaneously to the optimal
configuration of quarks~\cite{Hor85_PRD31,Wat89_NPA494,Hor91_PRC44,
Hor92_NPA536}. The string-flip model is thus reminiscent of the
Born-Oppenheimer approximation of wide use in atomic physics where
the fast degrees of freedom (electrons or gluons) adjust
instantaneously to changes in the slow degrees of freedom (ions and
constituent quarks).  The cornerstone of the string-flip model is a
many-body potential that: a) is symmetric under the exchange of
identical quarks, b) confines quarks within color-singlet clusters,
and c) enables these clusters to separate without generating
unphysical (long-range) van der Waals forces~\cite{Gre81_NPA370}.  
We regard these as a minimal characterization of any realistic
quark-based model of hadronic matter. To our knowledge, there is no
``conventional'' potential, namely, one that may be written as a sum
of two-body (or even three-body) forces, that satisfies the above
requirements. Thus, the string-flip potential is truly many body;
moving a single quark may affect the interaction among {\it all} the
quarks in the system. The many-body potential is obtained by solving a
difficult optimization problem~\cite{Mur80_S184}, as one must decide 
how to assign colored quarks into color-singlet clusters. This 
``quark-assignment'' problem is meant to represent the optimal
configuration of gluonic strings.

The manuscript has been organized as follows. In Sec.~\ref{formalism}
we define the string-flip potential and discuss some of its most
important characteristics. Further, we introduce a one-parameter
variational wave function that, in spite of its simplicity, is exact
in the low- and high-density limits. Indeed, at low density the
variational wave function favors a strong clustering of quarks into
color-singlet objects (``baryons''), while in the high-density limit
only Pauli correlations remain. Finally, medium modifications to the
properties of a heavy meson are addressed via a density-dependent {\it
heavy-quark} potential. In Sec.~\ref{results} we present our results
paying special emphasis to the correlations between the medium
modifications to the properties of the $J/\psi$-meson and the
deconfining phase transition. A summary and conclusions are presented
in Sec.~\ref{conclusions}.

\section{Formalism}
\label{formalism}

An exact many-body treatment of QCD at finite baryon density remains
an unsolved problem. Although fresh new ideas continue to emerge, we
use a {\it QCD-inspired model} to study the screening of color charge
by the nuclear medium. The main feature of the string-flip model
employed here is an {\it adiabatic} many-body potential characterized
by the arrangement of the gluonic strings (or flux-tubes) into a
minimum-energy configuration~\cite{Len86_AP170,Hor85_PRD31}.  Such a
many-body potential is symmetric under the exchange of identical
quarks and is able to confine quarks within color-singlet clusters
(``baryons'').  Yet the confining force saturates within a baryon,
thereby allowing the clusters to separate without generating
unobserved long-range van der Waals forces~\cite{Gre81_NPA370}.  While
the exact functional form of the potential remains uncertain, the
simultaneous requirements of quark confinement and cluster
separability are likely to require solving some type of {\it
quark-assignment} problem: which quarks belong to which color-singlet
cluster? In what follows, the most important features of the
string-flip model will be addressed. A more thorough discussion may be
found in Ref.~\cite{Tol02_PRC65}.

\subsection{The string-flip potential}
\label{string-flip}

Following a choice introduced in earlier 
references~\cite{Hor92_NPA536,Tol02_PRC65}, one starts by 
defining the optimal pairing between red and blue quarks as 
follows:
\begin{equation}
  V_{RB} = \mathop{\min}_{P}
           \sum_{i=1}^{A}
           v\Big({\bf r}_{iR},{\bf r}_{jB}\Big) \;,
 \label{vrb}
\end{equation}
where ${\bf r}_{iR}$ denotes the spatial coordinate of the {\it ith}
red quark and ${\bf r}_{jB}$ is the coordinate of its blue-quark 
``partner'' (${iB}\mapsto P({iB})\!\equiv\!{jB}$). Note that the 
minimization procedure is over all possible $A!$ permutations of 
the $A$ blue quarks and that the confining potential $v$ is assumed 
harmonic with a spring constant denoted by $k$. That is,
\begin{equation}
  v({\bf r}_{iR},{\bf r}_{jB}) = \frac{1}{2}k
  ({\bf r}_{iR}-{\bf r}_{jB})^{2} \;.
 \label{vstring}
\end{equation}
The ``blue-green'' and ``green-red'' components of the many-quark
potential are defined in direct analogy to Eq.~(\ref{vrb})
\begin{subequations}
\begin{eqnarray}
  V_{BG} &=& \mathop{\min}_{P}
             \sum_{i=1}^{A}
             v\Big({\bf r}_{iB},{\bf r}_{jG}\Big) \;,
  \label{vbg} \\
  V_{GR} &=& \mathop{\min}_{P}
             \sum_{i=1}^{A}
             v\Big({\bf r}_{iG},{\bf r}_{jR}\Big) \;.
  \label{vgr}
\end{eqnarray}
\end{subequations}
In this way, the many-body potential of the system is obtained
by simply adding the three different pairwise contributions
\begin{equation}
  V({\bf r}_{1},\ldots,{\bf r}_{N})=V_{RB}+V_{BG}+V_{GR} \;.
 \label{MQpotential}
\end{equation}
The many-body Hamiltonian describing a system of $N$ quarks each
with mass $m$ and momentum ${\bf p}_i$ is now given by
\begin{equation}
 H = \sum_{i=1}^{N}\frac{{\bf p}_{i}^{2}}{2m}+
     V({\bf r}_{1},\ldots,{\bf r}_{N}) \;,
 \label{HamiltonV}
\end{equation}

Note that the many-body potential defined in Eq.~(\ref{MQpotential})
is {\it flavor blind}. That is, the optimal pairing among the quarks
is done according to their color but not their flavor. In practice,
one may imagine collecting all quarks into red and blue ``buckets'' 
according to their color---but irrespective of their flavor---and 
then searching for their optimal pairing. This is then repeated 
for blue/green and green/red quarks. In the course of this work
we adopt units in which $k\!=\!m\!=\!1$.

\subsection{The variational wave function}
\label{wavefunction}

The complicated many-body dynamics may be captured with a simple
one-parameter variational wave function of the following form:
\begin{equation}
 \Psi_{\lambda}({\bf r}_{1},\ldots,{\bf r}_{N})=
  e^{-\lambda V({\bf r}_{1},\ldots,{\bf r}_{N})}
      \Phi_{FG}({\bf r}_{1},\ldots,{\bf r}_{N}) \;.
 \label{Psivar}
\end{equation}
Here $\lambda$ is the sole variational parameter, $V$ is the
many-body potential defined in Eq.~(\ref{MQpotential}), and
$\Phi_{FG}$ is a Fermi-gas wave function consisting of a product of
Slater determinants.  The variational parameter may be regarded as the
{\it order parameter} for the nuclear-to-quark-matter transition, as
$\lambda^{-1/2}$ represents the length-scale for quark
confinement. Indeed, at low density the average inter-quark separation
is much larger than the confining scale ($\lambda^{-1/2}\simeq 1)$ and
the clustering of three quarks into color-singlet clusters ({\it
nucleons}) ensues.  Note that while the interactions between quarks
within a single nucleon are strong, the residual nucleon-nucleon
interaction is weak, as the color force saturates within each
individual nucleon. This many-body feature of the model precludes the
development of long-range van der Waals forces.  As the density
increases, the average inter-quark separation will become comparable
to the confining scale. This will signal the transition to the quark
Fermi-gas phase. In the high-density regime the interactions between
quarks are weak ({\it asymptotic freedom}) and the system
``dissolves'' into a free Fermi gas of quarks. No correlations between
quarks remain, except those induced by the Pauli exclusion principle. 
Further details on the variational wave function may be found in 
Ref.~\cite{Tol02_PRC65}. For reference, the value of the variational
parameter for an isolated nucleon is given by 
$\lambda_{0}\!=\!1/\sqrt{3}$

\subsection{Variational Monte Carlo}
\label{MonteCarlo}

The structure of the variational wave function entails an important 
simplification, as the expectation value of the kinetic energy, which 
involves derivatives, may be simplified through an integration by
parts. This yields~\cite{Tol02_PRC65},
\begin{equation}
  \langle\Psi_{\lambda}|T|\Psi_{\lambda}\rangle \equiv
  \langle T \rangle_{\lambda}=
  T_{FG} + 2\lambda^2 \langle W \rangle_{\lambda} \;.
 \label{Tlambda}
\end{equation}
Here $T_{FG}$ is the kinetic energy of a ($\lambda\!=\!0$) free Fermi 
gas and $\langle W \rangle_{\lambda}$ is given by
\begin{equation}
    W  = \sum_{n=1}^{N}\frac{1}{m}
    \Big({\bf x}_{n}-{\bf y}_{n}\Big)^{2} \;,
 \label{Wpotent}
\end{equation}
where the sum is over all quarks in the system and ${\bf y}_{n}$
represents the average position of the two quarks connected to
the $n_{\rm th}$ quark located at ${\bf x}_{n}$. The increase in 
the kinetic energy of the system relative to that of a free Fermi 
gas is the result of clustering correlations. Of course, this 
``penalty'' is more than compensated by the potential energy which 
favors the clustering of quarks into nucleons at low densities. 
The expectation value of the total energy of the system now reduces 
to the following expression:
\begin{equation}
  E(\lambda) = T_{FG} + 2\lambda^2
               \langle W \rangle_{\lambda}
             + \langle V \rangle_{\lambda} \;.
 \label{ELambda2}
\end{equation}
This form is particularly convenient because the two functions 
that remain to be evaluated ($V$ and $W$) are local and may 
therefore be computed using standard Monte-Carlo 
techniques~\cite{Met53_JCP21}.

\subsection{The heavy-quark potential}
\label{heavyquark}

In this section we introduce the {\it heavy-quark} potential 
to correlate the modifications of the in-medium properties of
the $J/\psi$-meson to the deconfinement phase transition.
The heavy-quark potential is constructed as follows. Consider the
original system of $N$ quarks described by a variational wave 
function as in Eq.~(\ref{Psivar}). We assume that the variational
parameter $\lambda$ has been fixed at its optimal value. After a 
large number of thermalization sweeps, one measures the expectation 
value of the potential energy [Eq.~(\ref{MQpotential})]
\begin{equation}
 \langle V\rangle_{A}=
         \lim_{M\rightarrow\infty}\frac{1}{M}\sum_{m=1}^{M}
         V({\bf r}_{1}^{(m)},\ldots,{\bf r}_{N}^{(m)})\;, 		  
 \label{Vaverage}
\end{equation}
where the $M$ configurations of quarks are distributed according to
the square of the variational wave function. At the same time that one
pauses the Monte Carlo procedure to measure the potential energy in
the $m_{\rm th}$ configuration, one introduces a heavy quark-antiquark
({\it e.g.,} $c\bar{c}$) pair separated by a fixed distance $r$. This
procedure is implemented in our simulations by adding a {\it triplet
of static} quarks to the system; red and blue quarks fixed at the
origin and a green quark fixed a distance $r$ away. The red and blue
quarks are fixed at the same location to simulate a source of {\it
anti-green} color. As the heavy quarks are assumed static, their
masses play no role in the simulations. Having added the three heavy
quarks into the system, one now evaluates the potential energy of the
$(A+1)$-body system by re-computing the optimal pairing of quarks into
hadrons.  (Note that the added red and blue heavy quarks are
constrained to pair with each other.) That is,
\begin{equation}
 \langle V\rangle_{A+1}(r)=
         \lim_{M\rightarrow\infty}\frac{1}{M}\sum_{m=1}^{M}
         V({\bf r}_{1}^{(m)},\ldots,{\bf r}_{N}^{(m)}; r)\;.
 \label{Vaverage2}
\end{equation}
The heavy-quark potential is defined as the difference between
the potential energy of the $(A+1)$-body system relative to that
of the $A$-body system:
\begin{equation}
 V_{Q\bar{Q}}(r)=\langle V\rangle_{A+1}(r)
                -\langle V\rangle_{A}\;.
 \label{Vheavy}
\end{equation}
Note that the heavy-quark potential is computed in the {\it sudden}
approximation. That is, the many-quark wave function
[Eq.~(\ref{Psivar})] (and thus the location of all the light quarks in
the system) is assumed to remain unchanged as the heavy quarks are
introduced into the system. Not so, however, the pairing. The presence
of the medium screens the interaction between the heavy quarks by
finding the {\it optimal} pairing of the $N\!=\!3(A+1)$ quarks into
hadrons. In this way, the potential energy between the $Q\bar{Q}$ 
pair gets modified relative to its free-space form because of the
screening. In general, $V_{Q\bar{Q}}(r)\!\ne\!kr^{2}$.

\subsection{The $J/\psi$ meson in free space}
\label{JPsiinFreeSpace}

To quantify how a $J/\psi$-meson (or a comparable heavy meson)
is modified by color screening, we review its free-space properties 
in a model with harmonic confinement. Consider a $J/\psi$-meson as 
a nonrelativistic system of a quark-antiquark ($c\bar{c}$) pair of 
mass $m_{c}$ interacting via a harmonic confining potential of spring 
constant $2k$. That is,
\begin{equation}
 H = \frac{{\bf p}_{1}^{2}}{2m_{c}} +
     \frac{{\bf p}_{2}^{2}}{2m_{c}} +
     \frac{1}{2}(2k)({\bf r}_{1}-{\bf r}_{2})^{2} \;.
 \label{Hamilton}
\end{equation}
Introducing center-of-mass and relative coordinates
\begin{equation}
  {\bf R} = \frac{1}{2}
  ({\bf r}_{1}+{\bf r}_{2}) \;, \quad
  {\bf r} = ({\bf r}_{1}-{\bf r}_{2}) \;, 
 \label{Jacobi}
\end{equation}
enables one to reduce the above Hamiltonian to the following simple
form:
\begin{equation}
  H = \frac{P^{2}}{4m_{c}} 
    + \left(\frac{p^{2}}{m_{c}}+k{\bf r}^2\right) \;.
 \label{HJacobi}
\end{equation}
Describing the motion of the system relative to its center of mass,
the ground-state wave function of the $J/\psi$-meson in the present
harmonic approximation becomes
\begin{equation}
 \Phi_{J/\psi}({\bf r}) = 
      \frac{e^{-r^{2}/2b^{2}}}{(\pi b^{2})^{3/4}} 
      |\chi_{c}\chi_{\bar{c}}\rangle_{1M} \;, \quad
      b\equiv(km_{c})^{-1/4} \;.
 \label{FreeWF}
\end{equation}
Here $b$ is the harmonic-oscillator length and $\chi$ represents 
a spin-1/2 spinor. Using the above wave function, the ground-state 
energy and mean-square radius of the $J/\psi$-meson {\it in free 
space} are given by
\begin{subequations}
 \begin{eqnarray}
  &&
  E_{J/\psi}=2m_{c}+\frac{3}{2}\sqrt{\frac{4k}{m_{c}}}
                    \longrightarrow 11.342\;,\\
  &&
  r_{J/\psi}^{2}=\frac{3}{8}\frac{1}{\sqrt{km_{c}}}
                 \longrightarrow 0.168\;.
 \end{eqnarray}
 \label{JPsiGround}
\end{subequations}
The mean-square radius has been obtained by evaluating the
ground-state expectation value of the operator $r^{2}/4$. Moreover,
the arrows in Eq.~(\ref{JPsiGround}) indicate the appropriate
numerical values for the energy and the mean-square radius in units in
which $k\!=\!1$ and $m_{c}/m\!=\!5$. Note that the mass of the charm
quark ($\simeq 1.5$~GeV) has been fixed at five times the value of the
mass of the light constituent quarks ($\simeq 0.3$~GeV). Our goal now
is to quantify how these {\it baseline} values change as the density
of the medium increases and, further, to correlate these changes to
the deconfining phase transition.

\section{Results}
\label{results}

Variational Monte-Carlo calculations of a large number of ground-state
observables, such as the energy, the two-body correlation function,
and the strangeness-per-baryon ratio have been reported in an earlier
publication~\cite{Tol02_PRC65}. Here we borrow values for the
optimized variational parameter to compute the heavy-quark potential
as indicated in Eq.~(\ref{Vheavy}). Results obtained from these
simulations are displayed in Fig.~{\ref{Fig1} for a variety of
densities around the deconfining transition density of $\rho_{\rm
c}\!\simeq\!0.09$. Recall that the density is given in {\it
dimensionless units} (for a conversion to physical units see
Ref.~\cite{Tol02_PRC65}). The symbols in Fig.~{\ref{Fig1} denote the
outcome from the Monte-Carlo simulations while the lines represent
analytic fits to the simulation data. While it is evident that
quantitative changes develop below the critical density as a result of
screening, no major qualitative changes emerge. For example, the
$Q\bar{Q}$ potential remains confining for all densities below the
critical density $\rho_{\rm c}$. However, at densities above
$\rho_{\rm c}$, a dramatic change in the shape of the potential
emerges. Indeed, the heavy-quark potential ceases to be confining and
a $Q\bar{Q}$ spectrum supporting both bound and {\it continuum} states
develops. This behavior is reminiscent of the {\it temperature} 
dependence of the heavy-quark potential displayed in Fig.~6 of 
Ref.~\cite{Kar01_NPA605}.

\vspace{0.25in}
\begin{figure}[ht]
\begin{center}
\includegraphics[width=3.25in,angle=0,clip=false]{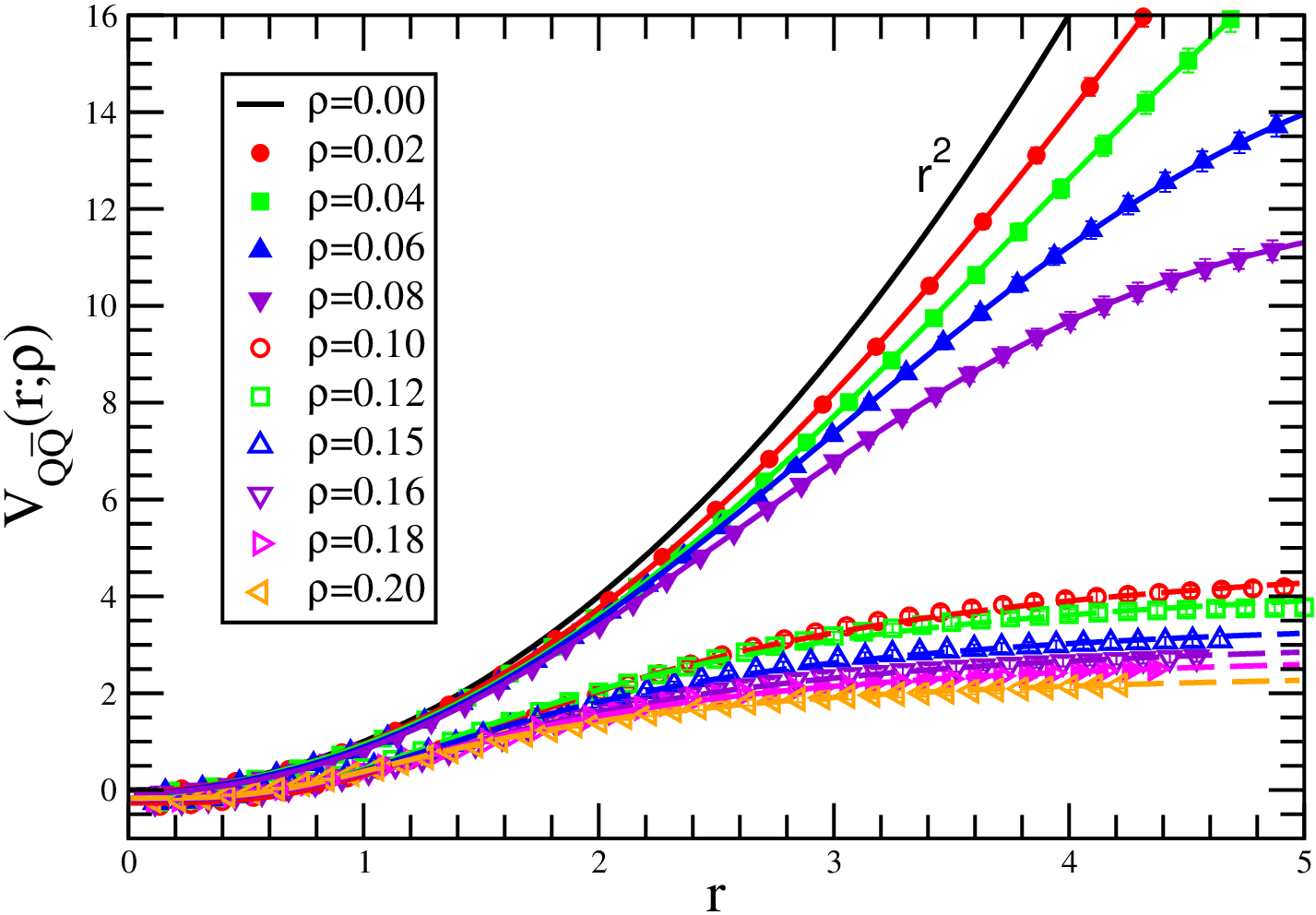}
\caption{Density dependence of the heavy-quark potential as a 
         function of the separation $r$ between the heavy-quark
         pair. The $\rho\!=\!0$ potential, 
	 $V_{Q\bar{Q}}(r)\!=\!r^2$, represents the baseline 
	 (free-space) potential.}
\label{Fig1}
\end{center}
\end{figure}

Medium modifications to the properties of the $J/\psi$ meson are
studied by solving Schr\"odinger's equation in the presence of the
heavy-quark potential. Recall that we have fixed the mass of the 
charm quark (in units of the light-quark constituent mass) to 
$m_{c}/m\!=\!5$. Expressing the ``reduced'' $J/\psi$ wave function 
as
\begin{equation}
 \Phi_{J/\psi}({\bf r}) = \frac{\varphi(r)}{r}
      |\chi_{c}\chi_{\bar{c}}\rangle_{1M} \;,
 \label{ScreenedWF}
\end{equation}
the Schr\"odinger equation for the ground-state ($l\!=\!0$) wave 
function takes the following simple form:
\begin{equation}
  \left(-\frac{\hbar^{2}}{2\mu_{c}}\frac{d^{2}}{dr^{2}}
   +V_{Q\bar{Q}}(r;\rho)\right)\varphi(r)=E\varphi(r) \;.
 \label{ScrhEq} 
 \end{equation}
Here $\mu_{c}\equiv m_{c}/2$ is the reduced mass of the $c\bar{c}$
pair. In Fig.~\ref{Fig2} we display the ground-state density
$\varphi^{2}(r)$ weighted by $r^{2}/4$, so that the area under the
curve yields directly the mean-square radius of the $J/\psi$-meson,
which we have enclosed in brackets. For densities below the
transition, the wave functions cluster around the free-space
value. Indeed, the change in the mean-square radius of the
$J/\psi$-meson between $\rho\!=\!0$ and $\rho\!=\!0.08$ amounts to
less than 6\%. However, as soon as the transition density is crossed,
a significant spread in the wave function develops, resulting in both
a reduced binding energy and an increased mean-square radius (see 
Fig.~\ref{Fig3}).  For comparison, at a density just above the 
transition ($\rho\!=\!0.10$), the energy decreases by almost 50\% 
and the mean-square radius increases by close to 25\%, relative to 
their free-space values.

\vspace{0.25in}
\begin{figure}[ht]
\begin{center}
\includegraphics[width=3.25in,angle=0,clip=false]{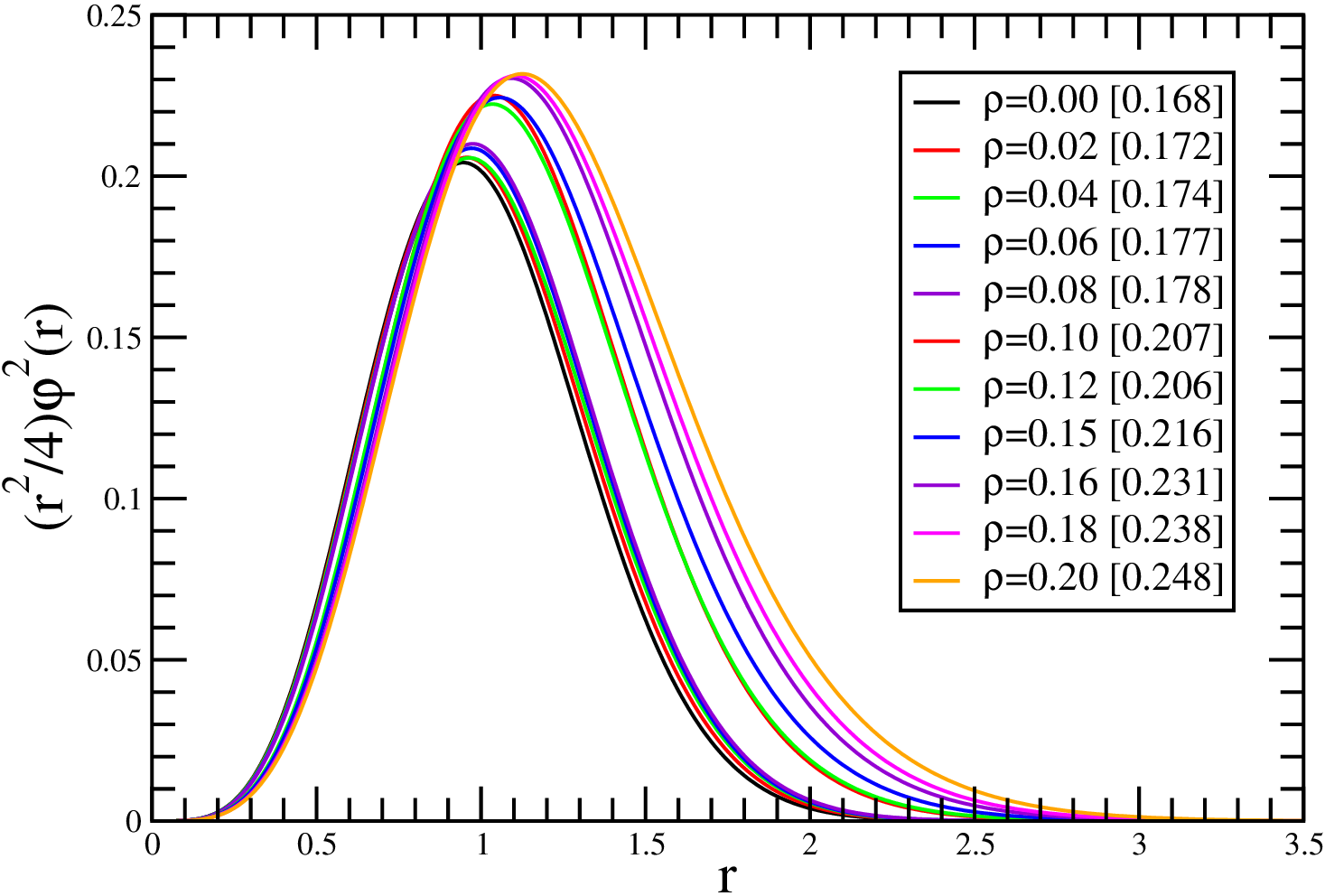}
\caption{Ground-state densities for the $J/\psi$ meson weighted
         by $r^{2}/4$ for a variety of densities. The quantities
         in brackets are the corresponding mean-square radii,
         which are obtained by integrating the area under the 
	 curves.}
\label{Fig2}
\end{center}
\end{figure}

Finally, we use Fig.~\ref{Fig3} to correlate modifications to the
properties of the $J/\psi$-meson in the medium to the deconfining
phase transition. Note that all shown quantities are given in units of
their free-space values.  The density dependence of the variational
parameter $\lambda$ exhibits an abrupt drop at a quark density of
about $\rho\!=\!0.09$. As $\lambda^{-1/2}$ represents the length-scale
for quark confinement, this drop signals the onset of the transition
to quark matter. Recall that for $\lambda\!\equiv\!0$ the only
remaining correlations in the many-quark wave function are due to the
Pauli-exclusion principle [see Eq.~(\ref{Psivar})]. While smaller in
magnitude, the corresponding abrupt changes in the energy and
mean-square radius of the $J/\psi$ are unmistakable.  This provides
confirmation of the long-time paradigm that modifications to the
properties of the $J/\psi$-meson in the medium represents a robust
indicator of the deconfining phase transition.  This in a model that,
while simple, uses exclusively quark degrees of freedom to describe
{\it dynamically} the deconfining phase transition. 

\vspace{0.35in}
\begin{figure}[ht]
\begin{center}
\includegraphics[width=2.8in,angle=0,clip=false]{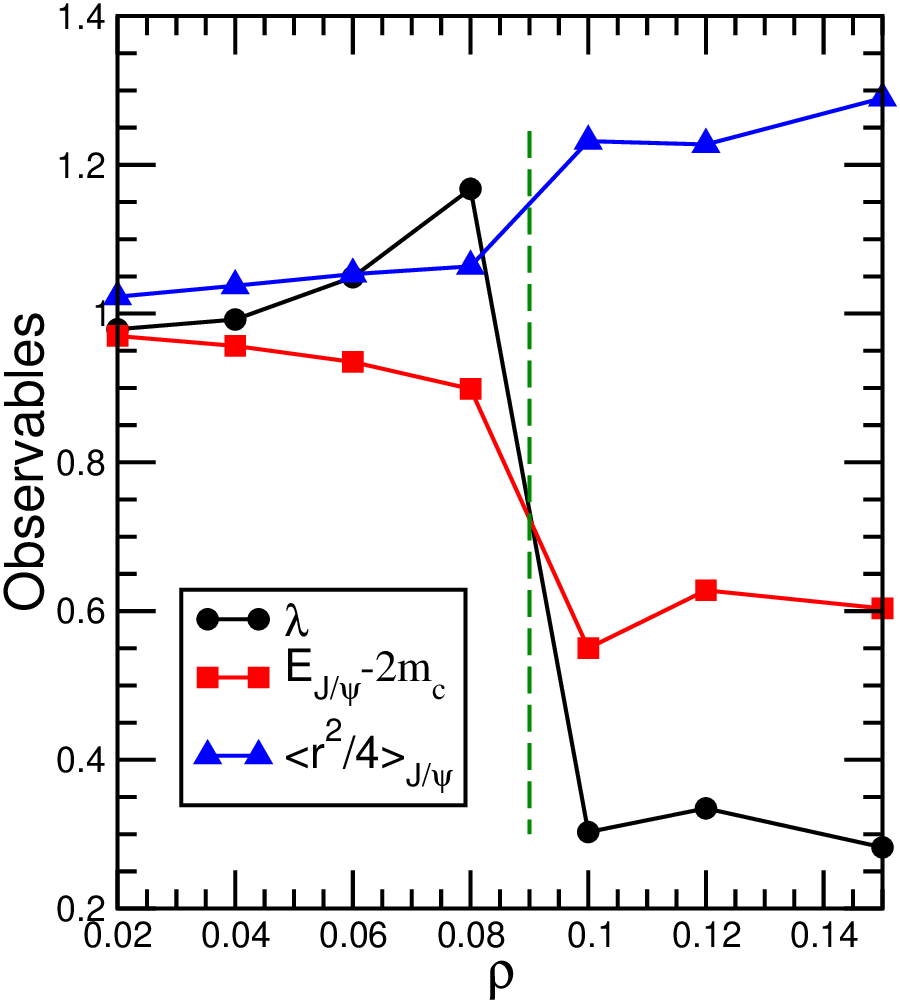}
\caption{Density dependence of the energy and mean-square
         radius of the $J/\psi$ meson. The ``jump'' in these
	 quantities at $\rho\!\simeq\!0.09$ is correlated to
	 the corresponding jump in the variational parameter
         $\lambda$ at the onset of the nuclear-to-quark matter 
	 transition. All quantities are measured in units of 
	 their free-space values.} 
\label{Fig3}
\end{center}
\end{figure}

\section{Conclusions}
\label{conclusions}

Modification to the properties of the $J/\psi$ meson in the nuclear
medium were computed in a constituent quark model of hadronic matter,
that confines quarks within individual color-neutral clusters, yet
allows the clusters to separate without generating unphysical
long-range forces. One of the greatest virtues of this {\it
string-flip} model is the description of the deconfining phase
transition. Indeed, the evolution of the system from a nuclear phase,
in which quarks clusters into color-singlet hadrons, to a quark
Fermi-gas phase is dynamical in the model; there is no need for
additional ad-hoc parameters to characterize the transition.

To quantify how the properties of the $J/\psi$ meson get modified by
color screening we computed the heavy-quark potential. Operationally,
this was effected by adding two heavy sources of color charge into the
system and monitoring how the interaction between the charges gets
modified by the presence of the medium. Screening of the heavy-quark
charges by the medium yields an effective potential that is
drastically different from its free-space value above the transition
density. Although the potential below the transition is also modified
by the medium, it retains its basic form. In contrast, above the
transition density the potential ceases to be confining and a ``heavy
quarkonium'' spectrum that supports both bound and continuum states
develops.

In our model one of the main indicators (``the order parameter'') of
the phase transition is the variational parameter of the many-quark
wave function. This parameter, which represents the length-scale for
quark confinement, drops abruptly to (almost) zero at the transition
density indicating how clustering correlations cease to be important
as the system evolves into the quark-matter phase. This drop
correlates strongly to similar abrupt changes in the in-medium
properties of the $J/\psi$-meson. This correlation confirms---in a
model that uses exclusively quarks and flux tubes (or strings) as the
fundamental degree of freedom---the long-standing tenet of Matsui and
Satz~\cite{Mat86_PLB178} of using medium modifications to the
$J/\psi$-meson as a robust signature for the onset of the deconfining
phase transition.

\acknowledgments
\medskip
We thank Prof. Anthony Frawley for many useful conversations. This 
work was supported in part by the United State Department of Energy 
under Contract No.DE-FG05-92ER40750 and by Conacyt M\'exico under 
grants 41600-A1, 41048-A1 and 42026-F.
\vfill

\bibliography{ColorScreening}
\end{document}